\documentclass[preprint,amsmath,amssymb,aps,showkeys,showpacs]{revtex4}
\usepackage[english]{babel}
\usepackage{graphicx}
\usepackage{graphics}
\usepackage{amsmath}
\usepackage{dcolumn}
\usepackage{amssymb}
\usepackage{bm}

\begin{document}
\title{A doubly anharmonic oscillator in an induced electric dipole system}
\author{K. Bakke}
\email{kbakke@fisica.ufpb.br}
\affiliation{Departamento de F\'isica, Universidade Federal da Para\'iba, Caixa Postal 5008, 58051-900, Jo\~ao Pessoa, PB, Brazil.}

\author{C. Salvador}
\email{charlie@fisica.ufpb.br}
\affiliation{Departamento de F\'isica, Universidade Federal da Para\'iba, Caixa Postal 5008, 58051-900, Jo\~ao Pessoa, PB, Brazil.}

\begin{abstract}

The behaviour of the interaction of the induced electric dipole moment of an atom with a uniform magnetic field and a non-uniform electric field are investigated in a rotating reference frame. An interesting aspect of this interaction is that it gives rise to an analogue of a spinless particle subject to the doubly anharmonic oscillator. Then, it is shown that analytical solutions to the Schr\"odinger equation can be obtained. Another point raised is that the quantum effects on the induced electric dipole moment can be observed if the uniform magnetic field possesses a discrete set of values.

\end{abstract}

%\pacs{03.65.Ge, 31.30.jc, 31.30.J-, 03.65.Vf}

\maketitle

\section{Introduction}

In recent decades, several studies have reported effects on quantum systems due to the interaction with a field configuration established by crossed magnetic and electric fields. Examples of these studies are the quasi-Landau behaviour in atomic systems \cite{cross11,cross12}, geometric quantum phases \cite{whw,whw2}, large electric dipole moments \cite{cross7} and systems of atoms and molecules \cite{cross1,cross2,cross3,cross4,cross5,cross6,cross8,cross9,cross10}. In particular, based on a field configuration of a uniform magnetic field and a radial electric field produced by a uniform volume distribution of electric charges inside a non-conductor cylinder, it has been shown in Ref. \cite{lin3} that the Landau levels of an atom with an induced electric dipole moment can be obtained. Recently, the field configuration proposal of Ref. \cite{lin3} has been extended to investigate effects in quantum rings \cite{dantas} and in rotating reference frames \cite{dantas1,ob3,ob5}. It is worth mentioning that this radial electric field produced by a uniform volume distribution of electric charges inside a non-conductor cylinder has been proposed in Ref. \cite{er} with the aim of discussing the quantum Hall effect for neutral particles with a permanent magnetic dipole moment. Further, this radial electric field has been used in studies of the scalar Aharonov-Bohm effect for a neutral particle with an electric quadrupole moment \cite{b} and the arising of a Coulomb-like potential in a neutral particle system with a magnetic quadrupole moment \cite{fb4}. 

Another point of view of the interaction between quantum systems and external fields is the presence of a non-uniform electric field. In Ref. \cite{b}, bound states solutions to a neutral particle with an electric quadrupole moment that interacts with an electric field produced by a non-uniform distribution of electric charges inside a non-conductor cylinder are discussed. In Refs. \cite{fb5,fb9}, Landau levels and rotating effects are investigated in a magnetic quadrupole moment system. In this work, we discuss the interaction of a neutral particle with an induced electric dipole moment with a field configuration given by a uniform magnetic field and a non-uniform electric field. In addition, we consider a rotating frame. We show that this interaction gives rise to an analogue of the doubly anharmonic oscillator \cite{doub3,heun}. By following Ref. \cite{doub3}, the doubly anharmonic oscillator is given by the scalar potential:
\begin{eqnarray}
V\left(r\right)=\varpi\,r^{2}+\lambda\,r^{4}+\eta\,r^{6},
\label{1}
\end{eqnarray}
where $\eta>0$ and $r=\sqrt{x^{2}+y^{2}}$ is the radial coordinate. In field theory, the doubly anharmonic oscillator has been dealt with the perturbation theory \cite{doub8,doub9,doub10,doub11,doub,doub1}. From these studies, analytical solutions to the Schr\"odinger equation for the doubly anharmonic oscillator have been investigated in Refs. \cite{doub2,doub3,doub4,doub5,doub7}.

The structure of this paper is: in section II, we introduce the quantum description of a neutral particle with an induced electric dipole moment in a rotating reference frame. We consider a field configuration given a uniform magnetic field and a non-uniform electric field. Then, we show that the interaction of the induced electric dipole moment of the neutral particle with this crossed magnetic and electric fields gives rise to an analogue of the doubly anharmonic oscillator potential (\ref{1}). Finally, we show that the Schr\"odinger equation can be solved analytically; in section III, we present our conclusions.

\section{Analogue of the doubly anharmonic oscillator}

In recent years, several works have analysed the effects of rotation on nonrelativistic quantum systems \cite{landau3,landau4,dantas1,anan,r13,fb5,ob3,ob5}. There, it is considered a rotating frame with a constant angular velocity given by $\vec{\Omega}=\Omega\,\hat{z}$. Then, it has been shown in Ref. \cite{dantas1} that the time-independent Schr\"odinger equation in this rotating frame is given by 
\begin{eqnarray}
\mathbb{H}_{0}\,\Psi-\vec{\Omega}\cdot\hat{L}\,\Psi=\mathcal{E}\Psi,
\label{1.1}
\end{eqnarray}
where $\mathbb{H}_{0}$ is the Hamiltonian operator of a particle system in the absence of rotation and $\hat{L}$ is the angular momentum operator.

In this work, our focus is on a nonrelativistic quantum system of a neutral particle (atom or molecule) with an induced electric dipole moment that interacts with external fields. By considering this quantum particle is moving with a velocity $v\ll c$, the quantum description of the interaction between the induced electric dipole moment with electric and magnetic fields is given by the following Hamiltonian operator \cite{whw,lin3,ob3}:
\begin{eqnarray}
\mathbb{H}_{0}=\frac{1}{2m}\left(\hat{p}+\alpha\,\vec{E}\times\vec{B}\right)^{2}-\frac{\alpha}{2}\,E^{2},
\label{1.2}
\end{eqnarray}
where $m=\bar{m}+\alpha\,B^{2}$, $\bar{m}$ is the mass of the neutral particle and we assume that $B^{2}=\mathrm{const}$. We shall work with units $\hbar=c=1$. Note in Eq. (\ref{1.2}) that $m$ is the mass of the particle, $\alpha$ is the dielectric polarizability, and $\vec{E}$ and $\vec{B}$ are the electric and magnetic fields in the laboratory frame, respectively. Besides, according to Ref. \cite{whw2}, the term $\alpha\,E^{2}$ given in Eq. (\ref{1.1}) is very small compared with the kinetic energy of the atoms, therefore we can neglect it without loss of generality from now on \footnote{A discussion about possible values of the dielectric polarizability in experimental contexts has been made in Ref. \cite{whw2}.}. From the Hamiltonian operator (\ref{1.1}), it is easy to observe that the angular momentum operator is given by $\hat{L}=\vec{r}\times\left(\hat{p}+\alpha\,\vec{E}\times\vec{B}\right)$.   

Henceforth, let us work with the cylindrical symmetry. In particular, by dealing with a planar system, we can write $\vec{r}=r\,\hat{r}$, where $r$ is the radial coordinate. Let us also consider a field configuration that interacts with the induced electric dipole moment of the atom given by the following magnetic and electric fields:
\begin{eqnarray}
\vec{B}=-B_{0}\,\hat{z};\,\,\,\vec{E}=\mu\,r^{3}\,\hat{r},
\label{1.3}
\end{eqnarray}
where $B_{0}$ is a constant and the electric field is given by a non-uniform electric charge density $u=\bar{\mu}\,r^{2}$ inside of a long non-conductor cylinder \cite{b}. Note that the magnetic field is in the $z$-direction, while the electric field is in the radial direction. By substituting Eqs. (\ref{1.2}) and (\ref{1.3}) into Eq. (\ref{1.1}), hence, the time-independent Schr\"odinger equation becomes
\begin{eqnarray}
\mathcal{E}\psi&=&-\frac{1}{2m}\nabla^{2}\psi-i\frac{\alpha\mu\,B_{0}}{m}\,r^{2}\,\frac{\partial\psi}{\partial\varphi}+\frac{\left(\alpha\mu\,B_{0}\right)^{2}}{2m}\,r^{6}\,\psi-i\Omega\,\frac{\partial\psi}{\partial\varphi}+\alpha\mu\,B_{0}\,\Omega\,r^{4}\psi,
\label{1.3a}
\end{eqnarray}
where $\nabla^{2}$ is the Laplacian in cylindrical coordinates. Let us write $\omega=\frac{\alpha\,\mu\,B_{0}}{m}$ as in Ref. \cite{lin3}. In addition, since the operators $\hat{L}_{z}=-i\,\partial_{\varphi}$ and $\hat{p}_{z}=-i\,\partial_{z}$ commute with the Hamiltonian operator given in the right-hand side of Eq. (\ref{1.3a}), therefore, we can write the solution to Eq. (\ref{1.3a}) in terms of their eigenvalues as $\psi\left(r,\,\varphi,\,z\right)=e^{il\varphi+ikz}\,R\left(r\right)$, where $l=0,\pm1,\pm2,\ldots$ and $-\infty\,<\,k\,<\,\infty$. In this way, we obtain the following radial equation:
\begin{eqnarray}
\frac{d^{2}R}{dr^{2}}+\frac{1}{r}\frac{dR}{dr}-\frac{l^{2}}{r^{2}}\,R-2m\omega\,l\,r^{2}\,R-m\omega\,\Omega\,r^{4}\,R-m^{2}\omega^{2}\,r^{6}\,R+\beta\,R=0,
\label{1.3b}
\end{eqnarray}
where $\beta=2m\left(\mathcal{E}-\Omega\,l\right)-k^{2}$. Note that we have in Eq. (\ref{1.3b}) an effective scalar potential given by $V_{\mathrm{eff}}=2m\omega\,l\,r^{2}+m\omega\,\Omega\,r^{4}+m^{2}\omega^{2}\,r^{6}$, which plays the role of the doubly anharmonic oscillator potential (\ref{1}). It stems from the interaction of the induced electric dipole moment of the neutral particle with the electric and magnetic fields (\ref{1.2}). Therefore, in the medium with the particular field configuration as given in Eq. (\ref{1.3}), the interaction of the induced electric dipole momentum with external fields gives rise to an analogue of the doubly anharmonic oscillator potential (\ref{1}). Next, let us define $x=\sqrt{\frac{m\omega}{2}}\,r^{2}$, and thus, Eq. (\ref{1.3b}) becomes
\begin{eqnarray}
\frac{d^{2}R}{dx^{2}}+\frac{1}{x}\frac{dR}{dx}-\frac{l^{2}}{4\,x^{2}}\,R-\lambda\,x\,R-x^{2}\,R+\frac{\bar{\beta}}{x}\,R-l\,R=0,
\label{1.4}
\end{eqnarray}
where we have defined the parameters:
\begin{eqnarray}
\lambda=\frac{\Omega}{m\omega}\sqrt{\frac{2}{m\omega}};\,\,\,\,\bar{\beta}=\frac{\beta}{\sqrt{2m\omega}}.
\label{1.5}
\end{eqnarray}

By analysing the asymptotic behaviour of Eq. (\ref{1.4}), i.e., the behaviour when $x\rightarrow\infty$ and $x\rightarrow0$, then, the solution to Eq. (\ref{1.4}) can be given in the form:
\begin{eqnarray}
R\left(x\right)=e^{-\frac{\lambda}{2}\,x}\,e^{-\frac{x^{2}}{2}}\,x^{\left|l\right|/2}\,H\left(x\right),
\label{1.6}
\end{eqnarray}
where $H\left(x\right)$ is a solution to the second order differential equation:
\begin{eqnarray}
\frac{d^{2}H}{dx^{2}}+\left[\frac{\left|l\right|+1}{x}-\lambda-2x\right]\frac{dH}{dx}+\left[\frac{\lambda^{2}}{4}-\left|l\right|-l-2-\frac{\lambda\left(1+\left|l\right|\right)-2\bar{\beta}}{2x}\right]H=0.
\label{1.7}
\end{eqnarray}
Note that Eq. (\ref{1.7}) is called in the literature as the biconfluent Heun equation \cite{heun}, and thus, the function $H\left(x\right)=H_{\mathrm{B}}\left(\left|l\right|,\,\lambda,\,\frac{\lambda^{2}}{4}-l,\,-2\bar{\beta};\,x\right)$ is the biconfluent Heun function.

Let us proceed with our analysis by using the Frobenius method \cite{arf,griff}. In this method, we write the function $H\left(x\right)$ as a power series around
the origin, i.e., $H\left(x\right)=\sum_{k=0}^{\infty}\,f_{k}\,x^{k}$. By substituting this series into Eq. (\ref{1.7}), we obtain two relations. The first relation is 
\begin{eqnarray}
f_{1}=\left[\frac{\lambda}{2}-\frac{\bar{\beta}}{1+\left|l\right|}\right]f_{0},
\label{1.8}
\end{eqnarray}
while the second relation corresponds to the recurrence relation:
\begin{eqnarray}
f_{k+2}=\frac{\left[\lambda\left(2k+\left|l\right|+3\right)-2\bar{\beta}\right]\,f_{k+1}-2\left(\frac{\lambda^{2}}{4}-\left|l\right|-l-2-2k\right)f_{k}}{2\left(k+2\right)\left(k+2+\left|l\right|\right)}.
\label{1.9}
\end{eqnarray}

Our objective is to find bound state solutions, then, we need that the function  $H\left(x\right)$ goes to zero when $x\rightarrow\infty$ and $x\rightarrow0$. For this reason, we must search for polynomial solutions to the biconfluent Heun equation (\ref{1.7}). Thereby, from the recurrence relation (\ref{1.9}), we have that the biconfluent Heun series terminates when we impose the following conditions:
\begin{eqnarray}
\frac{\lambda^{2}}{4}-\left|l\right|-l-2=2n;\,\,\,\,\,\,\,\,f_{n+1}=0,
\label{1.10}
\end{eqnarray}
with $n=1,2,3,\ldots$ being the quantum number associated with the radial modes. As an example, let us construct a polynomial of first degree ($n=1$) to $H\left(x\right)$. With $n=1$, the condition $\frac{\lambda^{2}}{4}-\left|l\right|-l-2=2n$ yields 
\begin{eqnarray}
\omega_{1,\,l}^{3}=\frac{\Omega^{2}}{2m^{3}\left(\left|l\right|+l+4\right)}.
\label{1.11}
\end{eqnarray}
Note that we have changed the notation of cyclotron frequency $\omega$ in Eq. (\ref{1.11}) and written it as $\omega=\omega_{n,\,l}$. This means that the cyclotron frequency can be adjusted with the purpose of achieving the polynomial of first degree to $H\left(x\right)$. Therefore, the allowed values of the cyclotron frequency that permit us to obtain a polynomial of first degree to $H\left(x\right)$ are given in Eq. (\ref{1.11}). Furthermore, since $\omega=\frac{\alpha\mu\,B_{0}}{m}$, then, we can extend the discussion made in Eq. (\ref{1.11}) to the magnetic field. Hence, from Eq. (\ref{1.11}), the discrete set of values of the magnetic field is given by
\begin{eqnarray}
B_{0}^{1,\,l}=\frac{m}{\alpha\,\mu}\left(\frac{\Omega^{2}}{2m^{3}\left[\left|l\right|+l+4\right]}\right)^{1/3}.
\label{1.11a}
\end{eqnarray}
It is worth pointing out that an analogous discussion about a discrete set of values of the magnetic field was made in Ref. \cite{furtado}. There, the magnetic field acquires a discrete set of values due to the effects of a topological defect and the self-interaction. Hence, our results agree with Ref. \cite{furtado}. 

We go further by analysing the condition $f_{n+1}=0$ given in Eq. (\ref{1.10}). For $n=1$, we have $f_{n+1}=f_{2}=0$. By using the recurrence relation (\ref{1.9}) and the relation (\ref{1.8}) to calculate the coefficient $f_{2}$, then, with $f_{n+1}=f_{2}=0$, we obtain the second degree algebraic equation for $\beta_{1,\,l}$:
\begin{eqnarray}
\beta_{1,\,l}^{2}-\frac{2\Omega}{m\omega_{1,\,l}}\left(2+\left|l\right|\right)\,\beta_{1,l}+\frac{\Omega^{2}}{m^{2}\omega_{1,\,l}^{2}}\left(3+\left|l\right|\right)\left(1+\left|l\right|\right)-4m\omega_{1,\,l}\,\left(1+\left|l\right|\right)=0.
\label{1.12}
\end{eqnarray}

Since we have defined $\beta=2m\left(\mathcal{E}-\Omega\,l\right)-k^{2}$, therefore, we obtain from Eq. (\ref{1.12}):
\begin{eqnarray}
\mathcal{E}_{1,\,l,\,k}=\Omega\,l+\frac{\Omega}{2m^{2}}\left(\frac{2m^{3}\left[\left|l\right|+l+4\right]}{\Omega^{2}}\right)^{1/3}\times\left\{2+\left|l\right|\pm\sqrt{1+\frac{2\left(1+\left|l\right|\right)}{\left(\left|l\right|+l+4\right)}}\right\}+\frac{k^{2}}{2m}.
\label{1.13}
\end{eqnarray}

Hence, Eq. (\ref{1.13}) give us the allowed energies associated with the radial mode $n=1$. Observe that the first term of  Eq. (\ref{1.13}) is the Page-Werner {\it et al} term \cite{r1,r2,r3}. It corresponds to the coupling between the angular momentum quantum number and the angular velocity. The last term corresponds to the free energy along the $z$-direction. Note that we reduce the system to a planar system by taking $k=0$. Further, by taking $\Omega\rightarrow0$, the effects of rotation vanish and there are no bound states solutions associated with the radial mode $n=1$. It is worth pointing out that it is possible to obtain the energy associated with other radial modes ($n=2,3,\ldots$) if we perform the same steps from Eq. (\ref{1.7}) to Eq. (\ref{1.13}).

\section{Conclusions}

In this work, we have investigated quantum effects on a neutral particle with an induced electric dipole moment in a rotating reference frame. Then, by analysing the interaction of the electric dipole moment with a uniform magnetic field and a non-uniform electric field, we have seen that an analogue of the doubly anharmonic oscillator potential (\ref{1}) arises from this interaction. Moreover, we have shown that the Schr\"odinger equation can be solved analytically. In particular, we have obtained the allowed energies associated with the radial mode $n=1$. These exact expressions for the allowed energies have been obtained from the analysis of the radial wave function, where we have searched for a polynomial of first degree to the biconfluent Heun series. As a consequence of constructing a polynomial of first degree to the biconfluent Heun function, we have seen that the magnetic field can have a discrete set of values (as we can see Eq. (\ref{1.11a})), otherwise, we cannot achieve this polynomial solution. Furthermore, we have seen a contribution to the allowed energies for the radial mode $n=1$ given by the coupling between the angular momentum quantum number and the angular velocity, which stems from the rotating effects \cite{r1,r2,r3}.

Despite the field configuration to be proposed in a theoretical point of view, it can be in the interests of the studies of atomic systems, since it opens new discussions about fields and quantum effects in elastic medium. As an example, an elastic medium with a topological defect \cite{kleinert,kat,tt7} can modify the electric field \cite{bf}. In the present case, the non-uniform electric field given in Eq. (\ref{1.3}) is hard to achieve it in the experimental context. An idea of achieving it is based on Ref. \cite{dop}. This non-uniform electric field could be made by successive processes of deposition of trivalent semiconductor materials, ring by ring and layer by layer, as the dopant inside the matrix of a tetravalent semiconductor material. It could be filled by increasing the charge from the symmetry axis to the edge of the cylinder, until the cylinder to be completed. If we consider a long non-conductor cylinder and each step of the deposition process to be sufficiently small, this non-continuous array of rings can produce a macroscopic effect, where non-uniform electric charge density inside the cylinder is proportional to $r^{2}$.

%\acknowledgments{The authors would like to thank the Brazilian agencies CNPq and CAPES for financial support.}

\section{Data accessibility statement}

This work does not have any experimental data.

\section{Authors' contributions}

K.B. and C.S. conceived the mathematical model, interpreted the results and wrote the paper. K.B. made most of the calculations in consultation with C.S. All authors gave final approval for publication.

\section{Competing interests}

We have no competing interests.

\section{Funding Statement}

K.B. would like to thank CNPq (grant number: 301385/2016-5) for financial support.

\section{Ethics statement}

This research poses no ethical considerations.

\end{document}